\documentclass[twocolumn,superscriptaddress,floatfix,showpacs]{revtex4}
\usepackage[latin9]{inputenc}
\usepackage{amsmath}
\usepackage{amssymb}
\usepackage{graphicx}

\makeatletter

\@ifundefined{textcolor}{}
{%
 \definecolor{BLACK}{gray}{0}
 \definecolor{WHITE}{gray}{1}
 \definecolor{RED}{rgb}{1,0,0}
 \definecolor{GREEN}{rgb}{0,1,0}
 \definecolor{BLUE}{rgb}{0,0,1}
 \definecolor{CYAN}{cmyk}{1,0,0,0}
 \definecolor{MAGENTA}{cmyk}{0,1,0,0}
 \definecolor{YELLOW}{cmyk}{0,0,1,0}
}

\@ifundefined{textcolor}{}{%
 \definecolor{BLACK}{gray}{0}
 \definecolor{WHITE}{gray}{1}
 \definecolor{RED}{rgb}{1,0,0}
 \definecolor{GREEN}{rgb}{0,1,0}
 \definecolor{BLUE}{rgb}{0,0,1}
 \definecolor{CYAN}{cmyk}{1,0,0,0}
 \definecolor{MAGENTA}{cmyk}{0,1,0,0}
 \definecolor{YELLOW}{cmyk}{0,0,1,0}
}

\makeatother

\begin{document}

\title{Nematic order in iron superconductors -- who is in the driver's seat?}

\author{R. M. Fernandes}

\affiliation{School of Physics and Astronomy, University of Minnesota, Minneapolis,
MN 55455, USA}

\author{A. V. Chubukov}

\affiliation{Department of Physics, University of Wisconsin-Madison, Madison,
Wisconsin 53706, USA}

\author{J. Schmalian}

\affiliation{Institut f\"ur Theorie der Kondensierten Materie, und Institut f\"ur
Festk\"orperphysik, Karlsruher Institut f\"ur Technologie, D-76131 Karlsruhe,
Germany}
\begin{abstract}
\textbf{Although the existence of nematic order in iron-based superconductors
is now a well-established experimental fact, its origin remains controversial.
Nematic order breaks the discrete lattice rotational symmetry by making
the $x$ and $y$ directions in the Fe plane non-equivalent. This
can happen because of (i) a tetragonal to orthorhombic structural
transition, (ii) a spontaneous breaking of an orbital symmetry, or
(iii) a spontaneous development of an Ising-type spin-nematic order
-- a magnetic state that breaks rotational symmetry but preserves
time-reversal symmetry. The Landau theory of phase transitions dictates
that the development of one of these orders should immediately induce
the other two, making the origin of nematicity a physics realization
of a ``chicken and egg problem''. The three scenarios are, however,
quite different from a microscopic perspective. While in the structural
scenario lattice vibrations (phonons) play the dominant role, in the
other two scenarios electronic correlations are responsible for the
nematic order. In this review, we argue that experimental and theoretical
evidence strongly points to the electronic rather than phononic mechanism,
placing the nematic order in the class of correlation-driven electronic
instabilities, like superconductivity and density-wave transitions.
We discuss different microscopic models for nematicity in the iron
pnictides, and link nematicity to other ordered states of the global
phase diagram of these materials -- magnetism and superconductivity.
In the magnetic model nematic order pre-empts stripe-type magnetic
order, and the same interaction which favors nematicity also gives
rise to an unconventional $s^{+-}$ superconductivity. In the charge/orbital
model magnetism appears as a secondary effect of ferro-orbital order,
and the interaction which favors nematicity gives rise to a conventional
$s^{++}$ superconductivity. We explain the existing data in terms
of the magnetic scenario, for which quantitative results have been
obtained theoretically, including the phase diagram, transport properties
of the nematic phase, scaling of nematic fluctuations, and the feedback
of the nematic order on magnetic and electronic spectra.} 
\end{abstract}
\maketitle

\section{Introduction}

The discovery of iron-based superconductors (FeSCs) with transition
temperatures $T_{c}$ as high as $65$K has signaled the beginning
of a new era in the investigation of unconventional superconductivity
(for a review, see \cite{review}). The key first step to unveil the
nature of the superconducting phase is to understand the normal state
from which superconductivity arises. In most FeSCs, superconductivity
is found in proximity to a magnetically ordered state (transition
temperature $T_{\mathrm{mag}}$), which led early on to the proposal
that magnetic fluctuations play the key role in promoting the superconducting
pairing \cite{review_Mazin,review_Andrey}. A more careful examination
of the phase diagram, however, revealed that there is another non-superconducting
ordered state besides magnetism. Namely, at a certain temperature
$T_{\mathrm{nem}}$, the system spontaneously breaks the symmetry
between the $x$ and $y$ directions in the Fe plane, reducing the
rotational point group symmetry of the lattice from tetragonal to
orthorhombic, while time-reversal symmetry is preserved. In some materials,
such as hole-doped (Ba$_{1-x}$K$_{x}$)Fe$_{2}$As$_{2}$, the tetragonal-to-orthorhombic
and magnetic transitions are simultaneous and first-order ($T_{\mathrm{nem}}=T_{\mathrm{mag}}$),
whereas in electron-doped Ba(Fe$_{1-x}$Co$_{x}$)$_{2}$As$_{2}$
and isovalent-doped BaFe$_{2}$(As$_{1-x}$P$_{x}$)$_{2}$, they
are split ($T_{\mathrm{nem}}>T_{\mathrm{mag}}$) and second order
\cite{Kim11,Birgeneau11,matsuda_t} (see Fig. 1). As doping increases,
the $T_{\mathrm{nem}}$ line tracks the $T_{\mathrm{mag}}$ line across
the phase diagram, approaching the superconducting dome. It is therefore
essential to understand the origin of this new order as it may support
or act detrimentally to superconductivity.

\begin{figure}
\begin{centering}
\includegraphics[width=1\columnwidth]{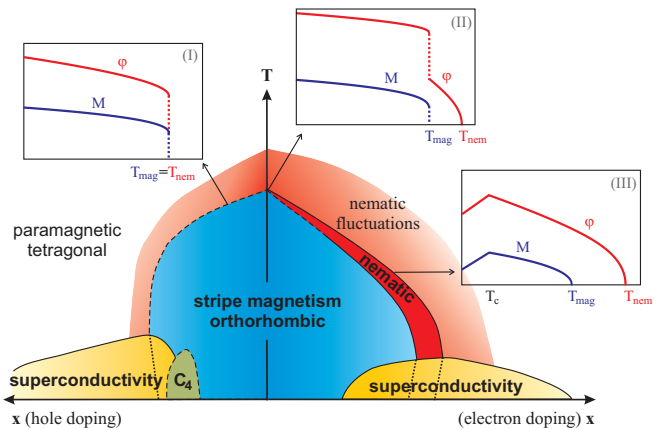} 
\par\end{centering}

\caption{Schematic phase diagram of hole-doped and electron-doped iron pnictides
of the BaFe$_{2}$As$_{2}$ family. The blue area denotes stripe-type
orthorhombic magnetism, the red area denotes nematic/orthorhombic
paramagnetic order, and the yellow area, superconductivity. The green
area corresponds to a magnetically-ordered state that preserves tetragonal
($C_{4}$) symmetry, as observed recently \cite{Osborn13}. The shaded
red region denotes a regime with strong nematic fluctuations. Bent-back
dotted lines illustrate the magnetic and nematic transition lines
inside the superconducting dome. Second-order (first-order) transitions
are denoted by solid (dashe) lines. The insets show the temperature-dependence
of the nematic ($\varphi$) and magnetic ($M$) order parameters in
different regions of the phase diagram: region (I) corresponds to
simultaneous first-order magnetic and nematic transitions; region
(II), to split second-order nematic and first-order magnetic transitions;
and region (III) to split second-order transitions. }
\end{figure}

The order parameter for a transition in which a rotational symmetry
is broken but time-reversal symmetry is preserved is a director (i.e.
a vector without an arrow), similar to the order parameter in the
nematic phase of liquid crystals \cite{Fradkin_review}. By analogy,
the orthorhombic state in FeSCs has been called a ``nematic state''.
Unlike isotropic liquid crystals, however, in FeSCs the lattice symmetry
forces the director to point only either along $x$ or $y$ directions,
what makes the nematic order parameter Ising-type (Ising-nematic).

At first sight, one might view this tetragonal-to-orthorhombic transition
as a regular structural transition driven by lattice vibrations (phonons).
However, experiments find anisotropies in several electronic properties,
such as the dc resistivity \cite{Chu10,Tanatar10}, to be much larger
than the anisotropy of the lattice parameters. This led to the idea
that the tetragonal-to-orthorhombic transition may be driven by electronic
rather than lattice degrees of freedom. If this is the case, then
the transition into the nematic phase is driven by the same fluctuations
that give rise to superconductivity and magnetic order, and therefore
is an integral part of a global phase diagram of FeSCs. Electronic
nematic phases have been recently proposed in other unconventional
superconductors, such as high-$T_{c}$ cuprates and heavy-fermion
materials~\cite{Fradkin_review}. An electronically driven nematic
state in FeSCs would be in line with a generic reasoning that the
pairing in all these correlated electron systems has the same origin.

The discussion on the ``nematicity'' in FeSCs has been largely focused
on two key issues: (i) Can the experiments distinguish ``beyond reasonable
doubt'' between phonon-driven and electron-driven tetragonal symmetry
breaking? (ii) If this transition is driven by electrons, which of
their collective degrees of freedom are driving it - charge/orbital
fluctuations or spin fluctuations? Answering the last question is
crucial for the understanding of superconductivity in FeSCs because
we argue below that charge/orbital fluctuations favor a sign-preserving
$s$-wave state ($s^{++}$) whereas spin fluctuations favor a sign-changing
$s$-wave ($s^{+-}$) or a $d$-wave state. Here we give our perspective
on these issues, discuss the phenomenology of the nematic state, its
experimental manifestations, and the underlying microscopic models.

\section{Phenomenology of the nematic phase}

To describe the nematic state, the first task is to identify the appropriate
order parameter. The experimental manifestations of nematic order
can be clustered into three classes. Taken alone, each class points
to a different origin of the nematic phase (see Fig. 2 for schematic
representation): 
\begin{itemize}
\item Structural distortion -- the lattice parameters $a$ and $b$ along
the $x$ and $y$ directions become different \cite{Kim11}. Such
an order is normally associated with a phonon-driven structural transition; 
\item Charge orbital order -- the occupations $n_{xz}$ and $n_{yz}$ (and
on-site energies) of the $d_{xz}$ and $d_{yz}$ Fe-orbitals become
different \cite{Yi2011}. The appearance of such an order is normally
associated with divergent charge fluctuations; 
\item Spin-nematic order -- the static spin susceptibility $\chi_{\mathrm{mag}}\left(\mathbf{q}\right)$
becomes different along the $q_{x}$ and $q_{y}$ directions of the
Brillouin zone before a conventional SDW state develops~\cite{matsuda_t}.
The appearance of such an order is normally associated with divergent
quadrupole magnetic fluctuations. 
\end{itemize}
The fact that these three order parameters are non-zero in the nematic
phase leads to a dilemma, which can be best characterized as the physics
realization of a ``chicken and egg problem'': all three types of
order (structural, orbital, and spin-nematic) must be present no matter
who drives the nematic instability. This follows from the fact that
bi-linear combinations of the order parameters which break the same
symmetry (in our case, the tetragonal symmetry of the system) are
invariant under symmetry transformations and must therefore appear
in the Landau free energy. Suppose that one of the three order parameters
is the primary one, i.e. its fluctuations drive the nematic instability.
Let's call it $\psi_{1}$ and the other two $\psi_{2}$ and $\psi_{3}$.
The free energy has the generic form 
\begin{eqnarray}
 &  & F\left[\psi_{1},\psi_{2},\psi_{3}\right]=\frac{1}{2}\chi_{1}^{-1}\psi_{1}^{2}+\frac{b}{4}\psi_{1}^{4}\nonumber \\
 &  & +\lambda_{12}\psi_{1}\psi_{2}+\frac{1}{2}\chi_{2}^{-1}\psi_{2}^{2}+\lambda_{13}\psi_{1}\psi_{3}+\frac{1}{2}\chi_{2}^{-1}\psi_{3}^{2}+...\label{GL}
\end{eqnarray}

\begin{figure}
\begin{centering}
\includegraphics[width=1\columnwidth]{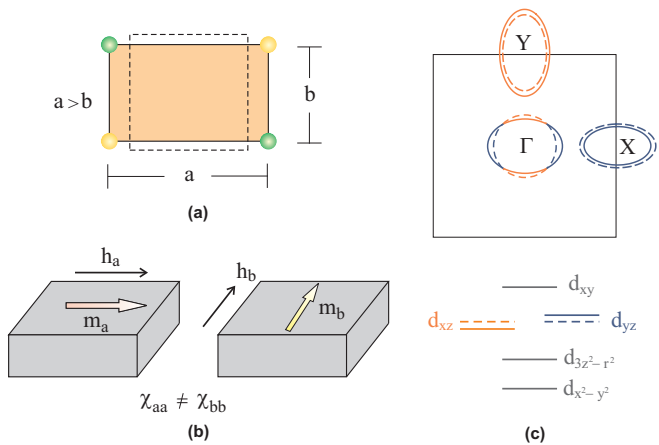} 
\par\end{centering}

\caption{Manifestations of nematic order in the iron pnictides: (a) Structural
distortion from a tetragonal (dashed line) to an orthorhombic (solid
line) unit cell \cite{Kim11}. (b) Anisotropy in the uniform spin
susceptibility $\chi_{ij}=m_{i}/h_{j}$, where $m_{i}$ denotes the
magnetization along the $i$ direction induced by a magnetic field
$h_{j}$ applied along the $j$ direction \cite{matsuda_t}. (c) Splitting
of the $d_{xz}$ and $d_{yz}$ orbitals (orange and blue lines, respectively)
\cite{Yi2011}. The corresponding distortion of the Fermi surface
is also shown (see also Fig. 5a).}
\end{figure}

Because the nematic transition is driven by $\psi_{1}$, the coefficient
$\chi_{1}$, which corresponds to the order parameter susceptibility
in the disordered state, diverges at $T=T_{\mathrm{nem}}$ and becomes
negative for $T<T_{\mathrm{nem}}$, while $\chi_{2}$ and $\chi_{3}$
remain finite and positive (although fluctuations of $\psi_{2}$ and
$\psi_{3}$ may shift slightly $T_{\mathrm{nem}}$). For $T<T_{\mathrm{nem}}$,
$\psi_{1}$ orders on its own: $\left\langle \psi_{1}\right\rangle =\pm\left(-\chi_{1}^{-1}/b\right)^{1/2}$.
If $\lambda_{ij}$ in (\ref{GL}) were zero, the other two fields
$\psi_{2}$ and $\psi_{3}$ would not order, but once $\lambda_{ij}$
are finite, a non-zero $\left\langle \psi_{1}\right\rangle $ instantly
induces finite values of the secondary order parameters $\left\langle \psi_{2}\right\rangle =-\lambda_{12}\chi_{2}\left\langle \psi_{1}\right\rangle $,
$\left\langle \psi_{3}\right\rangle =-\lambda_{13}\chi_{3}\left\langle \psi_{1}\right\rangle $.
As a consequence, there is only one nematic transition temperature
at which all three $\left\langle \psi_{i}\right\rangle $ become non-zero
(e.g., lattice symmetry is broken at the same temperature where electronic
nematic order emerges), and it is not possible to determine who causes
the instability by looking solely at equilibrium order parameters.
An additional experimental complication is the presence of nematic
twin domains below $T_{\mathrm{nem}}$, what effectively averages
$\left\langle \psi_{1}\right\rangle $ to zero. This problem can be
circumvented by applying a small detwinning uniaxial stress \cite{Chu10,Tanatar10},
which acts as a conjugate field to $\psi_{1}$ and breaks the tetragonal
symmetry at all temperatures, making $T_{\mathrm{nem}}$ an ill-defined
quantity.

One way to select the primary order is to carefully study fluctuations
in the symmetry-unbroken phase at $T>T_{\mathrm{nem}}$. Because the
primary order parameter $\psi_{1}$ acts as an external field for
the secondary order parameters, $\psi_{2}$ and $\psi_{3}$, fluctuations
of the former renormalize the susceptibilities of the latter to 
\begin{equation}
\tilde{\chi}_{2}\approx\chi_{2}\left(1+\lambda_{12}^{2}\chi_{2}\chi_{1}\right),\tilde{\chi}_{3}\approx\chi_{3}\left(1+\lambda_{13}^{2}\chi_{2}\chi_{1}\right),\label{susceptibility}
\end{equation}
 where $\chi_{1}=\left\langle \psi_{1}^{2}\right\rangle $ is the
susceptibility of the primary field. The renormalized susceptibilities
of the secondary fields do diverge at the nematic transition, however
for small enough $\lambda_{12}$ and $\lambda_{13}$, $\tilde{\chi}_{2}$
and $\tilde{\chi}_{3}$ begin to grow only in the immediate vicinity
of $T_{\mathrm{nem}}$, where $\chi_{1}$ is already large. If one
can measure the three susceptibilities independently, Eq. (\ref{susceptibility})
in principle provides a criterion to decide which order parameter
drives the instability. The implementation of this procedure is possible
(see next section), but is complicated by two factors. First, it only
works if $\lambda_{12}$ and $\lambda_{13}$ are relatively weak,
what normally implies that the systems falls into the weak/moderate
coupling category. If the coupling is large, all three order parameters
become so inter-connected that the question ``who is in the driver's
seat?'' becomes meaningless. Second, in some FeSCs the nematic transition
is first order, in which case all three susceptibilities jump from
one finite value to another, even before the susceptibility of the
primary field gets enhanced.

\section{Experimental evidence for electronic nematicity}

\subsection{Measurements in the nematic phase}

The first evidence for the electronic character of the tetragonal-to-orthorhombic
transition came from resistivity measurements in detwinned samples
\cite{Chu10,Tanatar10}, which revealed that resistivity anisotropies
are significantly larger than relative lattice distortions and also
display a nontrivial dependence on doping and disorder \cite{Nakajima2011}.
Other non-equilibrium quantities, such as thermopower \cite{Jiang2013}
and optical conductivity\cite{Dusza2011,Nakajima2011}, were also
found to display large anisotropies, which in optical measurements
were observed to extend to energies of several hundreds of $\mathrm{meV}$.
Anisotropies in observables related to charge and spin were also seen:
angle-resolved photoemission spectroscopy (ARPES) found a splitting
between the on-site energies of the $d_{xz}$ and $d_{yz}$ orbitals,
indicative of ferro-orbital order \cite{Yi2011} and torque magnetometry
revealed different uniform magnetic susceptibilities along the $x$
and $y$ directions \cite{matsuda_t}. The onset of magnetic anisotropy
coincides with the observation of a non-zero orthorhombic distortion,
in agreement with the discussions of the previous section. Strong
signatures of emerging magnetic anisotropy were also found in the
behavior of the nuclear magnetic resonance (NMR) lines across $T_{\mathrm{nem}}$
\cite{Imai12}.

The direct observation of electronic anisotropy in the nematic state
was made possible by scanning tunneling microscopy (STM). The first
measurements, performed deep inside the magnetic phase, found that
the local density of states around an impurity is characterized by
a dimer-like structure extended along the magnetic ordering vector
direction \cite{Davis10}. Subsequent measurements showed that these
dimers persist above $T_{\mathrm{mag}}$, in the temperature regime
of the nematic state \cite{Rosenthal13}. An additional piece of evidence
in favor of the electronic character of the nematic transition came,
ironically, from x-ray measurements of the orthorhombic distortion
inside the SC phase. These measurements found a strong suppression
of the distortion below $T_{c}$ \cite{Nandi10}, what is a characteristic
signature of the competition for the same electronic states between
two electronically-driven orders.

\subsection{Measurements in the tetragonal phase}

A few recent measurements focused on fluctuations in the tetragonal
state, in particular, on the shear modulus $C_{s}$, which is the
inverse susceptibility of the structural order parameter \cite{Fernandes2010,Yoshizawa2012,Boehmer2013}.
If the structural transition is driven not by the lattice but by some
other electronic degree of freedom, Eq. (\ref{susceptibility}) provides
a natural way to connect the shear modulus to the electronic nematic
susceptibility $\chi_{1}$. An experimentally observed softening of
the shear modulus above $T_{\mathrm{nem}}$ was successfully fitted
by Eq. (\ref{susceptibility}) using both magnetic \cite{Fernandes2010}
and charge/orbital \cite{Yoshizawa2012} phenomenological models for
$\chi_{1}$, indicating that structural distortion is very likely
not the primary order.

Perhaps the strongest evidence that the nematic transition is electronically-driven
came from the recent measurements of the anisotropy of the resistivity
\cite{Chu2012}. Using a piezoelectric, the measurements were performed
by using strain (the structural distortion) as the control parameter,
rather than stress, as in previous setups. The strain $\delta$ is
one of the order parameter fields in the free energy Eq. (\ref{GL}).
Using the resistivity anisotropy $\rho_{\mathrm{anis}}=\rho_{xx}-\rho_{yy}$
as a proxy of the nematic order parameter, it was experimentally shown
that the susceptibility $\partial\rho_{\mathrm{anis}}/\partial\delta$
diverges near the nematic transition. This is only possible if the
structural distortion is a conjugate field to the primary order parameter,
rather than the primary order parameter itself -- otherwise $\rho_{\mathrm{anis}}$
would be simply proportional to the order parameter $\delta$, with
a constant prefactor.

\section{Microscopic models for electronic nematicity}

A successful microscopic theory for electronic nematicity must describe
the global phase diagram of FeSCs, i.e. not only the nematic order
but also magnetism and superconductivity. A popular starting point
is the multi-orbital Hubbard model, which describes hopping between
all Fe-As orbitals and local interactions, such as intra-band and
inter-band Hubbard repulsions and Hund's exchange \cite{review_Mazin}.
There is a general agreement among researches that this model does
contain all information about the phase diagram. The model has been
analyzed at both weak/intermediate coupling, when the system is a
metal, and at strong coupling, when electrons on at least some orbitals
were assumed to be localized or ``almost localized''. The nematic
order has been obtained in both limits, what is yet another indication
that it is a generic property of FeSCs. We adopt the itinerant approach,
since most FeSCs are metals. In this itinerant scenario, the low-energy
electronic states lie around hole-like Fermi-surface pockets at the
center of the Fe-square lattice Brillouin zone and electron-like Fermi-surface
pockets at the borders of the Brillouin zone, see Fig. 5a. The microscopic
reasoning for either magnetic or orbital scenarios of electronic nematicity
follows from two different assumptions about the sign of the effective
inter-pocket interaction $U$ \cite{Eremin10}, which is a combination
of the Hubbard and Hund interactions dressed up by coherence factors
associated with the transformation from the orbital to the band basis.
As we will see, each scenario leads to a prediction of a particular
superconducting pairing state.

\begin{figure}
\begin{centering}
\includegraphics[width=1\columnwidth]{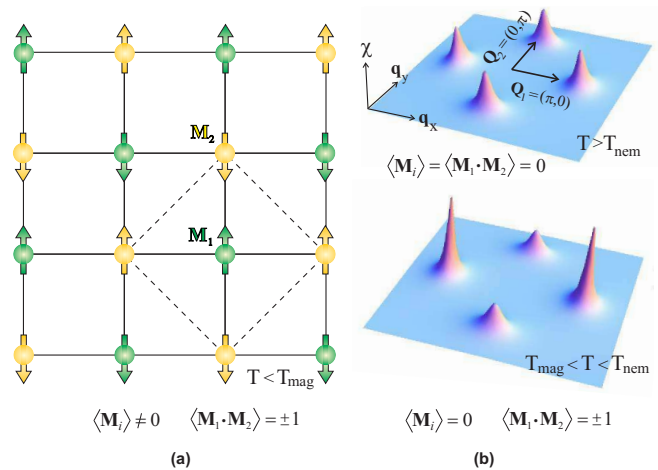} 
\par\end{centering}

\caption{Nematic order in both real and momentum space. In (a), we show the
stripe magnetic configuration in real space, which can be interpreted
as two inter-penetrating Neel sublattices (green and yellow) with
staggered magnetization $\mathbf{M}_{1}$ and $\mathbf{M}_{2}$. In
terms of the two magnetic order parameters $\mathbf{M}_{X}$ and $\mathbf{M}_{Y}$
defined in momentum space and used throughout the text, we have $\mathbf{M}_{1,2}=\mathbf{M}_{X}\pm\mathbf{M}_{Y}$.
In (b), we show the onset of nematic order in the paramagnetic phase
($\left\langle \mathbf{M}_{i}\right\rangle =0$), in terms of the
magnetic susceptibility $\chi\left(\mathbf{q}\right)$ across the
first Brillouin zone. For $T>T_{\mathrm{nem}}$, the two inelastic
peaks at $\mathbf{Q}_{X}=\left(\pi,0\right)$ and $\mathbf{Q}_{Y}=\left(0,\pi\right)$
have equal amplitudes, i.e. $\left\langle M_{X}^{2}-M_{Y}^{2}\right\rangle \equiv\left\langle \mathbf{M}_{1}\cdot\mathbf{M}_{2}\right\rangle =0$.
For $T_{\mathrm{mag}}<T<T_{\mathrm{nem}}$, one of the peaks becomes
stronger than the other, i.e. $\left\langle M_{X}^{2}-M_{Y}^{2}\right\rangle \equiv\left\langle \mathbf{M}_{1}\cdot\mathbf{M}_{2}\right\rangle \neq0$,
which breaks the equivalence between the $x$ and $y$ directions. }
\end{figure}

\subsection{Magnetic scenario}

The magnetic mechanism for the nematic order follows from the observation
that in most FeSCs the observed magnetic order on the Fe atoms is
of stripe type, with ordering vectors $\mathbf{Q}_{X}=\left(\pi,0\right)$
or $\mathbf{Q}_{Y}=\left(0,\pi\right)$ i.e. spins are parallel to
each other along one direction and anti-parallel along the other \cite{Dai_review}
(see Fig. 3a). This order breaks not only the $O(3)$ spin-rotational
symmetry (and time-reversal symmetry), but it also breaks the $90^{\circ}$
lattice rotational symmetry down to $180^{\circ}$ by choosing the
ordering vectors to be either $\mathbf{Q}_{X}$ or $\mathbf{Q}_{Y}$.
This additional tetragonal symmetry breaking enhances the order parameter
manifold to $O(3)\times Z_{2}$ \cite{Sachdev,Kivelson}. In terms
of the two magnetic order parameters $\mathbf{M}_{X}=\sum_{k}c_{{\bf k}+{\bf Q}_{X},\alpha}^{\dagger}{\bf \boldsymbol{\sigma}}_{\alpha\beta}c_{{\bf k},\beta}$
and $\mathbf{M}_{Y}=\sum_{k}c_{{\bf k}+{\bf Q}_{Y},\alpha}^{\dagger}{\bf \boldsymbol{\sigma}}_{\alpha\beta}c_{{\bf k},\beta}$,
associated with the ordering vectors $\mathbf{Q}_{X}$ and $\mathbf{Q}_{Y}$,
the breaking of the $O(3)$ symmetry implies $\left\langle \mathbf{M}_{i}\right\rangle \neq0$
while the breaking of the $Z_{2}$ symmetry implies $\left\langle M_{X}^{2}\right\rangle \neq\left\langle M_{Y}^{2}\right\rangle $
\cite{Fernandes12}. In a mean-field approach both $O(3)$ and $Z_{2}$
symmetries are broken simultaneously at $T_{\mathrm{mag}}$. However,
fluctuations split the two transitions and give rise to an intermediate
phase at $T_{\mathrm{mag}}<T<T_{\mathrm{nem}}$ where tetragonal symmetry
is broken but the spin-rotational $O(3)$ symmetry is not, i.e. $\left\langle M_{X}^{2}\right\rangle \neq\left\langle M_{Y}^{2}\right\rangle $
while $\left\langle \mathbf{M}_{i}\right\rangle =0$. This is by definition
a nematic order, which, viewed this way, is an unconventional magnetic
order which preserves time-reversal symmetry (a spin nematic). In
real space, the stripe magnetic state can be viewed as two inter-penetrating
Neel sublattices with staggered magnetizations $\mathbf{M}_{1}=\mathbf{M}_{X}+\mathbf{M}_{Y}$
and $\mathbf{M}_{2}=\mathbf{M}_{X}-\mathbf{M}_{Y}$. In terms of these
quantities, the nematic state is characterized by $\left\langle \mathbf{M}_{1}\cdot\mathbf{M}_{2}\right\rangle \neq0$
while $\left\langle \mathbf{M}_{i}\right\rangle =0$ (see Fig. 3).

Within a microscopic description, the instability towards a stripe
magnetic order is associated with the divergence of the static spin
susceptibility $\chi_{\mathrm{mag}}\left(\mathbf{Q}\right)$. Without
any interactions, the bare particle-hole susceptibility $\chi_{0}\left(\mathbf{Q}\right)$
is by itself sizable at $\mathbf{Q}_{X}$ and $\mathbf{Q}_{Y}$ because
these wave-vectors connect electronic states at the hole and electron
pockets. When the inter-pocket interaction $U$ is positive (repulsive),
there is an additional RPA-type enhancement of the spin susceptibility,
roughly as $\chi_{\mathrm{mag}}\left(\mathbf{Q}\right)=\chi_{0}\left(\mathbf{Q}\right)/\left(1-U\chi_{0}\left(\mathbf{Q}\right)\right)$,
and at some $T=T_{\mathrm{mag}}$, $\chi_{\mathrm{mag}}\left(\mathbf{Q}_{X,Y}\right)$
diverges. This however does not guarantee that the magnetic order
is of stripe type as the latter emerges only if below $T_{\mathrm{mag}}$
$\left\langle \mathbf{M}_{X}\right\rangle \neq0$ and $\left\langle \mathbf{M}_{Y}\right\rangle =0$,
or vise versa. To determine which magnetic state develops, one needs
to calculate higher order terms in the magnetic free energy \cite{Fernandes12,Eremin10,Brydon11}.
The result is that at least for small dopings, the system selects
the stripe order. The static nematic susceptibility $\chi_{\mathrm{nem}}$
(the correlator of $M_{X}^{2}-M_{Y}^{2}$) can be obtained by including
fluctuations of the nematic order parameter $M_{X}^{2}-M_{Y}^{2}$,
yielding: 
\begin{equation}
\chi_{\mathrm{nem}}=\frac{T\sum\limits _{\mathbf{Q}}\chi_{\mathrm{mag}}^{2}\left(\mathbf{Q}\right)}{1-g\, T\sum\limits _{\mathbf{Q}}\chi_{\mathrm{mag}}^{2}\left(\mathbf{Q}\right)}\label{eq_nematics}
\end{equation}
 where $T$ is the temperature, and $g\propto U^{2}$ is the composite
coupling which, when positive, sets the magnetic order to be of stripe
type. In dimensions $d<4$, $\sum\limits _{\mathbf{q}}\chi_{\mathrm{mag}}^{2}\left(\mathbf{q}\right)$
diverges at $T_{\mathrm{mag}}$ (assuming that the magnetic transition
is second order). Eq. (\ref{eq_nematics}) then shows that the nematic
susceptibility diverges at a higher $T_{\mathrm{nem}}>T_{\mathrm{mag}}$,
when $g\sum\limits _{\mathbf{q}}\chi_{\mathrm{mag}}^{2}\left(\mathbf{q}\right)=1$,
i.e. at sufficiently large but still finite magnetic correlation length.
This mechanism naturally ties the nematic and magnetic ordering temperatures
to each other over the entire phase diagram. In between $T_{\mathrm{nem}}$
and $T_{\mathrm{mag}}$, the $Z_{2}$ symmetry is broken but $O(3)$
is not, i.e., $\left\langle M_{X}^{2}\right\rangle \neq\left\langle M_{Y}^{2}\right\rangle $
but $\left\langle \mathbf{M}_{i}\right\rangle =0$. The difference
between $T_{\mathrm{nem}}$ and $T_{\mathrm{mag}}$ is stronger in
quasi-2D systems where $T_{\mathrm{mag}}$ is further decreased by
thermal fluctuations, while $T_{\mathrm{nem}}$ remains unaffected
\cite{Kivelson,batista}.

More detailed microscopic calculations show that for some system parameters
the nematic transition is second order, but for other input parameters
it becomes first-order \cite{Fernandes12}. In the latter case, a
jump in the nematic order parameter induces a jump in the magnetic
correlation length, which may instantaneously trigger a first-order
magnetic transition. In any case, when the Fermi pockets are decomposed
into their orbital characters, one finds within the same microscopic
model that the emergence of spin-nematic order gives rise to orbital
order $\Delta n=n_{xz}-n_{yz}$, since the electron pocket at $ $$\mathbf{Q}_{X}$
has mostly $d_{yz}$ character, whereas the electron pocket at $\mathbf{Q}_{Y}$
has mostly $d_{xz}$ character. Similarly, a spin-nematic order induces
a structural distortion $a\neq b$ \cite{Qi09,Cano10}.

We see therefore that the repulsive inter-pocket interaction $U>0$
enhances spin fluctuations, which gives rise to both magnetism and
nematicity. To describe the global phase diagram of FeSCs, one needs
also to investigate superconductivity. Spin fluctuations peaked at
${\bf Q}_{X}$ and ${\bf Q}_{Y}$ strongly enhance inter-pocket repulsion,
which becomes larger than intra-pocket repulsion. In this situation,
the system is known to develop either an unconventional $s^{+-}$
superconductivity, in which the gap functions have different signs
in the hole and in the electron pockets, or a $d_{x^{2}-y^{2}}$ superconductivity
\cite{review_Mazin,review_Andrey}. We emphasize that spin-nematic
order and $s^{+-}$ superconductivity are both intrinsic consequences
of the same magnetic scenario.

Other microscopic models also find nematic order in proximity to a
magnetic instability. For instance, explicit evaluation of the ferro-orbital
susceptibility using the multi-orbital Hubbard model finds that it
is enhanced only in the presence of spin fluctuations \cite{Kontani12},
similarly to what is described by Eq. (\ref{eq_nematics}). Studies
of models with both localized and itinerant orbitals also found \cite{Phillips11,Dagotto13,w_ku10}
that the proximity to magnetism is an important ingredient for orbital
order. In purely localized-spin models the interplay between magnetism
and ferro-orbital order is blurred by the complicated form of the
effective Hamiltonian, which deviates from a simpler Kugel-Khomskii
type \cite{kruger09,Applegate11}.

\subsection{Charge/orbital scenario}

In its simplest form, the charge/orbital scenario for the nematic
order parallels the magnetic scenario, the only difference being the
sign of the interaction $U$ between electron and hole pockets. If
this interaction turns out to be negative, it is the charge/orbital
susceptibility rather than the spin susceptibility that is enhanced
as $\chi_{\mathrm{orb}}\left(\mathbf{Q}\right)=\chi_{0}\left(\mathbf{Q}\right)/\left(1+U\chi_{0}\left(\mathbf{Q}\right)\right)$,
diverging at ${\bf Q}_{X}$ and ${\bf Q}_{Y}$ at a certain $T_{\mathrm{orb}}$.
This divergence would signal the onset of a charge density-wave state
with ordering vectors $\mathbf{Q}_{X}$ or $\mathbf{Q}_{Y}$ (or both)
and order parameters $W_{X}=\sum_{k}c_{{\bf k}+{\bf Q}_{X},\alpha}^{\dagger}\delta_{\alpha\beta}c_{{\bf k},\beta}$
and $W_{Y}=\sum_{k}c_{{\bf k}+{\bf Q}_{Y},\alpha}^{\dagger}\delta_{\alpha\beta}c_{{\bf k},\beta}$.
This order breaks translational symmetry and, like in the magnetic
scenario, breaks also an additional $Z_{2}$ symmetry if only one
order parameter becomes non-zero. It is natural to expect, although
no explicit calculations have been done to the best of our knowledge,
that fluctuations split the temperatures at which the translational
and the $Z_{2}$ symmetries are broken, in a manner similar to Eq.
(\ref{eq_nematics}). Then, in the intermediate temperature range
$T_{\mathrm{orb}}<T<T_{\mathrm{nem}}$, the system spontaneously develops
ferro-orbital order in which $\left\langle W_{X}^{2}\right\rangle \neq\left\langle W_{Y}^{2}\right\rangle $
while $\left\langle W_{X}\right\rangle =\left\langle W_{Y}\right\rangle =0$.
A structural distortion and the difference between $\left\langle M_{X}^{2}\right\rangle $
and $\left\langle M_{Y}^{2}\right\rangle $ appear instantly once
ferro-orbital order sets in. However, magnetic order only appears
at a smaller temperature, presumably via changes in the magnetic correlation
length induced by the ferro-orbital order.

For the Cooper pairing, the orbital scenario implies that the inter-pocket
interaction is attractive and enhanced. Once this interaction exceeds
the intra-pocket repulsion, the system develops a superconducting
instability towards an $s^{++}$ state -- a conventional pairing state
where the gap functions have the same sign in all pockets \cite{review_Andrey}.

What we described above is the simplest scenario for orbital order.
More complex models have been also proposed to account for the nematic
transition without involving magnetic degrees of freedom. In Ref.
\cite{Littlewood}, it was proposed that nematicity could arise as
an unequal hybridization between localized $d_{xy}$ orbitals and
itinerant $d_{xz}/d_{yz}$ orbitals. In Ref. \cite{Tesanovic11} it
was suggested that both spin and charge interactions are present and
that the larger interaction in the spin channel gives rise to magnetic
order at, say, $\mathbf{Q}_{X}$. However, before this happens, a
weaker charge interaction gives rise to charge order at the other
momentum $\mathbf{Q}_{Y}$ (a pocket density-wave state), which would
break the tetragonal symmetry of the system. Whether such a pocket
density-wave is experimentally realized in FeSCs remains to be seen.

\section{Comparison between theory and experiment}

\begin{figure}
\begin{centering}
\includegraphics[width=1\columnwidth]{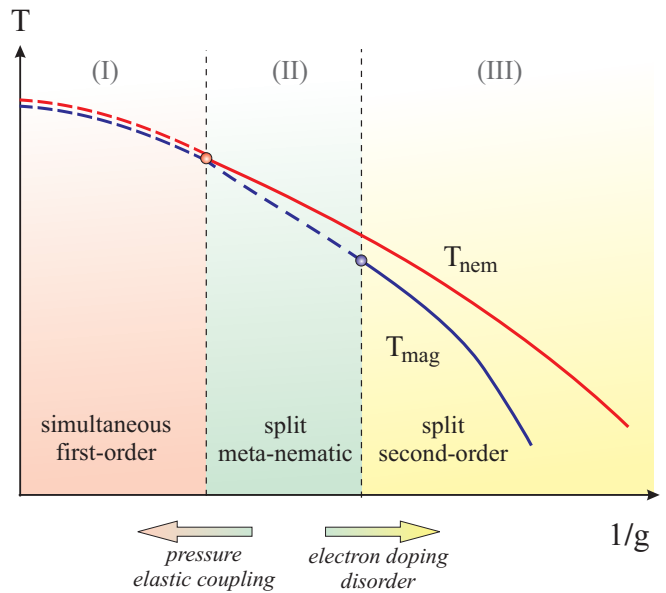} 
\par\end{centering}

\caption{Schematic representation of the evolution of the magnetic and nematic
transitions as function of the inverse nematic coupling $g$, according
to the microscopic itinerant spin-nematic model. Second-order (first-order)
lines are denotes by solid (dashed) lines. Regions (I)-(III) correspond
to those of the phase diagram in Fig. 1. The arrows show how the nematic
order parameter $g$ is expected to change as function of various
control parameters.}
\end{figure}

Although the experimental evidence presented in Section III favors
an electronic nematic instability, disentangling the orbital and magnetic
scenarios is difficult on a qualitative level, what begs for a more
direct comparison between microscopic models and experimental results.
In this regard, the doping evolution of the magnetic and structural
transitions is an important benchmark. BaFe$_{2}$As$_{2}$, one of
the compounds most extensively investigated, displays a second-order
nematic transition at $T_{\mathrm{nem}}$ followed by a a first-order
``meta-nematic transition'' at a lower $T$, where the system simultaneously
undergoes a first-order magnetic transition. The meta-nematic transition
has been observed by x-ray \cite{Kim11,Birgeneau11} and torque magnetometry
\cite{matsuda_t}, although the data disagree on the precise value
of $T_{\mathrm{nem}}$. As charge carriers are introduced in the system
via Co substitution in the Fe sites, the splitting between the two
transitions increases, and eventually the meta-nematic transition
disappers and the magnetic transition becomes second-order.

How does this compare to theory? For the magnetic scenario, a detailed
theoretical analysis \cite{Fernandes12} shows that three types of
system behavior are possible in systems that are moderately anisotropic,
depending on the value of the nematic coupling $g$ (see Fig. 4).
At large $g$, nematic and magnetic transitions are simultaneous and
first order. At intermediate $g$, nematic order develops via a second-order
transition, and there is a meta-nematic transition at a lower $T$,
where magnetic order also develops discontinuously. At smaller $g$,
nematic and magnetic transitions are separate and second-order, with
an intermediate spin-nematic phase between $T_{\mathrm{nem}}$ and
$T_{\mathrm{mag}}$. The microscopic calculations found that $g$
decreases with electron doping, and the theoretical phase diagram
in Fig. 4 is fully consistent with the one for the electron-doped
Ba(Fe$_{1-x}$Co$_{x}$)$_{2}$As$_{2}$ if we place the $x=0$ point
in the region II in Fig. 4. No calculations of how the nematic and
magnetic transitions evolve with carrier concentration have been done
within the charge/orbital scenario.

One can take the comparison with the data even further and compare
the two versions of the magnetic scenario -- for itinerant and for
localized spins. Both predict stripe magnetic order and pre-emptive
$Z_{2}$ symmetry-breaking but differ in the details. In particular,
in localized models $g$ is generally small and is unaffected by carrier
concentration \cite{Chandra90,Kivelson}. This makes the description
of the doping dependence in the localized spin approach somewhat problematic,
although not impossible~\cite{batista}. A more essential difference
is that in localized models the coupling $g$ is always positive,
while in itinerant models $g$ may become negative at large enough
hole doping \cite{Lorenzana08,Brydon11}. For negative $g$, there
is no tetragonal symmetry breaking either above or below the magnetic
transition as the system selects a tetragonally-symmetric combination
of both $\mathbf{Q}_{X}$ and $\mathbf{Q}_{Y}$ orders. A symmetry-preserving
magnetic state with orders at $\mathbf{Q}_{X}$ and $\mathbf{Q}_{Y}$
has been recently observed in Ba$_{1-x}$Na$_{x}$Fe$_{2}$As$_{2}$
\cite{Osborn13} and Ba(Fe$_{1-x}$Mn$_{x}$)$_{2}$As$_{2}$ \cite{Kim10}
at large enough doping -- a strong argument in favor of the itinerant
magnetic scenario.

Another key quantity to compare experiment and theory is the resistivity
anisotropy. Deep in the magnetically ordered state, the anisotropic
folding of the Fermi surface plays the major role in determining the
resistivity anisotropy \cite{Tanatar10,bascones}. In the nematic
state, $T>T_{\mathrm{mag}}$, orbital order and spin-nematic order
have different effects on the dc resistivity: while the former causes
an anisotropy in the Drude weight \cite{Devereaux10,Phillips11},
the latter gives rise to anisotropy in the scattering rate \cite{Fernandes11}.
The calculated anisotropy in the Drude weight has the opposite sign
to the one observed experimentally \cite{Devereaux10,Phillips11},
whereas the calculated anisotropy in the magnetic scattering rate
was shown to agree with experiments, including a sign-change of the
anisotropy between electron-doped and hole-doped materials \cite{Blomberg13}.

\begin{figure}
\begin{centering}
\includegraphics[width=1\columnwidth]{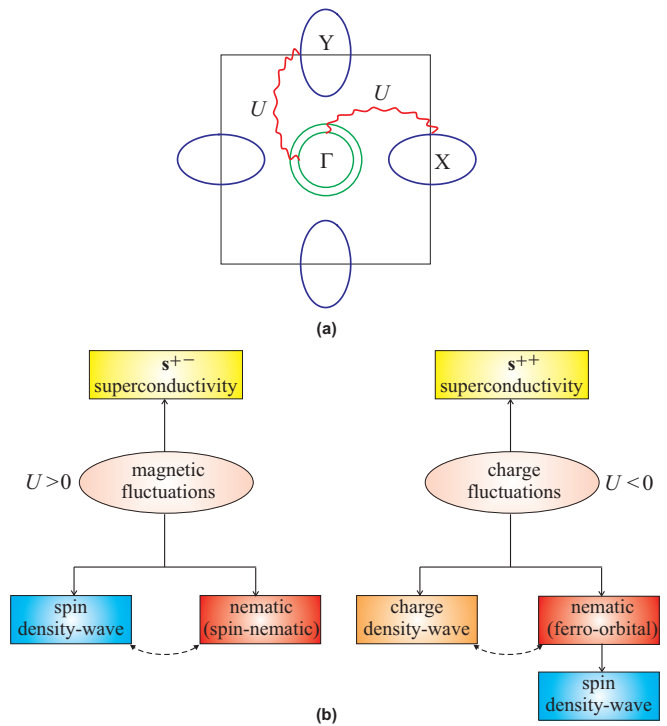} 
\par\end{centering}

\caption{(a) The minimal Fermi surface with hole pockets (green lines) at the
center $\Gamma$ of the Fe-square lattice Brillouin zone and electron
pockets (blue lines) centered at the $X=\left(\pi,0\right)$ and $Y=\left(0,\pi\right)$
points of the Brillouin zone. $U$ is the inter-pocket interaction
discussed in the main text. (b) Depending on the sign of $U$, either
spin fluctuations ($U>0$, repulsion) or charge fluctuations ($U<0$,
attraction) dominate. In the former, a stripe-type spin density-wave
state is pre-empted by a spin-nematic phase, and the superconducting
state is $s^{+-}$ (opposite-sign gaps around the hole and electron
pockets). In the latter, a stripe-type charge density-wave is pre-empted
by a charge-nematic phase, and the superconducting state is $s^{++}$
(same-sign gaps around the hole and electron pockets). In this scenario,
magnetic order only appears as a secondary consequence of ferro-orbital
order.}
\end{figure}

One can also compare theoretical and experimental results for the
feedback effects from the nematic order on the electronic and the
magnetic spectrum ~\cite{Phillips11,Fernandes12,Goswami11}. In the
magnetic scenario, nematic order enhances the magnetic correlation
length, what gives rise to strong magnetic fluctuations and a possible
pseudogap in the electronic spectrum. A significant increase of magnetic
fluctuations below $T_{\mathrm{nem}}$ has been observed via NMR in
compounds where $T_{\mathrm{nem}}$ and $T_{\mathrm{mag}}$ are well
separated \cite{Ma_NaFeAs}. Also, recent ARPES experiments found
the pseudogap behavior (a suppression in the density of states at
low energies) whose onset coincides with the nematic transition \cite{pseudogap_Matsuda}.
Within the orbital scenario, the key feedback from the orbital order
is a Pomeranchuk distortion of the Fermi surface induced by orbital
order \cite{Phillips11}. 

Nematic fluctuations above $T_{\mathrm{nem}}$ have also been used
to compare experiment and theory. Orbital fluctuations have been argued
to affect the density of states at the Fermi level\cite{Phillips12}
and leave signatures in point-contact spectroscopy consistent with
the data \cite{Greene12}. Alternatively, one can employ Eq. (\ref{susceptibility})
to compare the renormalized lattice susceptibility (the shear modulus
$C_{s}$), assumed to be non-critical, with the susceptibility $\chi_{1}$
associated with either the orbital or the spin-nematic order parameter.
Eq. (\ref{susceptibility}) must be satisfied if the corresponding
electronic order drives the nematic instability. In Ref. \cite{Gallais13},
a quasi-elastic peak in the Raman response was attributed to charge/orbital
fluctuations and used to extract the corresponding orbital susceptibility.
On the other hand, the spin-nematic susceptibility, being proportional
to $\sum_{\mathbf{q}}\chi_{\mathrm{mag}}^{2}\left(\mathbf{q}\right)$
(see Eq. \ref{eq_nematics}), can be measured via the NMR spin-lattice
relaxation rate $1/T_{1}$. Comparison with shear modulus data for
a family of electron-doped FeSCs found that there is a robust scaling
between $C_{s}$ and $1/T_{1}$ data \cite{Fernandes13_shear}. This
provides strong support to the idea that the nematic transition is
magnetically-driven.

\section{Perspectives}

The bulk of experimental and theoretical results which we presented
in this mini-review supports the idea that nematic order in FeSCs
is of electronic origin, what places it at par with other known electronic
instabilities such as superconductivity or density-wave orders. It
is likely that magnetic fluctuations drive the nematic instability.
In any case, all three orders -- spin-nematic, orbital, and structural,
appear simultaneously below $T_{\mathrm{nem}}$. The important question
not addressed until very recently is the role of nematicity for high-temperature
superconductivity. It is unlikely that nematic fluctuations can mediate
superconductivity as spin or charge fluctuations do, but nematic fluctuations
may nevertheless enhance $T_{c}$ by reducing the bare intra-pocket
repulsion. Below $T_{c}$, however, nematic order has been found to
compete with superconductivity \cite{Nandi10,Moon12}, like density-wave
orders do. A special case in which nematicity strongly affects $T_{c}$
is when $s$-wave and $d$-wave superconducting instabilities are
nearly degenerate, what was suggested to be the case for strongly
hole-doped and strongly electron-doped FeSCs \cite{Taillefer12}.
In this situation nematic order leads to a sizeable enhancement of
$T_{c}$ by lifting the frustration associated with the competing
pairing states \cite{Livanas12,Fernandes_Millis13,DungHaiLee13}.
These results clearly point to the need of additional investigations
of the interplay between nematicity and superconductivity.

\section{Acknowledgement}

We acknowledge useful discussions with E. Abrahams, J. Analytis, E.
Bascones, A. Böhmer, J. van den Brink, P. Brydon, S. Bud'ko, P. Canfield,
P. Chandra, P. Dai, M. Daghofer, L. Degiorgi, I. Eremin, I. Fisher,
Y. Gallais, A. Goldman, A. Kaminski, V. Keppens, D. Khalyavin, M.
Khodas, S. Kivelson, J. Knolle, H. Kontani, A. Kreyssig, F. Krüger,
W. Ku, W.C. Lee, J. Lorenzana, W. Lv, S. Maiti, D. Mandrus, R. McQueeney,
Y. Matsuda, I. Mazin, C. Meingast, A. Millis, R. Osborn, A. Pasupathy,
I. Paul, P. Phillips, R. Prozorov, S. Sachdev, Q. Si, T. Shibauchi,
L. Taillefer, M. Takigawa, M. Tanatar, M. Vavilov, P. Wölfle, and
M. Yoshizawa. The authors benefited a lot from the discussions with
our great colleague Z. Tesanovic who unexpectedly passed away last
year. A.V.C. is supported by the Office of Basic Energy Sciences U.S.
Department of Energy under the grant \#DE-FG02-ER46900.

\end{document}